\documentstyle[12pt]{article}
\begin{document}
\baselineskip=16pt
\bibliographystyle{unsrt}
\newcommand{\beq}{\begin{equation}}
\newcommand{\eeq}{\end{equation}}
\newcommand{\bea}{\begin{eqnarray}}
\newcommand{\eea}{\end{eqnarray}}
\newcommand{\etal}{{\it et al.}}
\newcommand{\sol}{M_\odot}
\newcommand{\be}{\begin{equation}}
\newcommand{\ee} {\end{equation}}
\newcommand{\ba}{\begin{eqnarray}}
\newcommand{\ea}{\end{eqnarray}}
  
\begin{center}
\begin{Large}
{\bf 
A Model for Non High Energy Gamma Ray Bursts and Sources of Ultra High
Energy Cosmic Rays\\
-super strongly magnetized milli-second pulsar formed from a (C+O) star  and a neutron star (black hole)  close binary system- 
} \end{Large} \vspace*{1cm}\\
 
Takashi NAKAMURA \\
Yukawa Institute for Theoretical Physics,
Kyoto University, 606, Japan \vspace*{0.5cm}\\

{\bf ABSTRACT}\\
\end{center}
 
{As a progenitor of NHE (Non High Energy) GRBs (Gamma Ray Bursts), we 
propose  a (C+O) star  and a neutron star (black hole)  close binary system 
with separation $\sim$ 0.2 R$_\odot$. Since the (C+O) star is tidally locked,
the collapsing core should have a spin angular momentum $\sim 6\times
10^{48}$cgs  so that a new born pulsar should be  a milli-second pulsar(MSP)
($P\sim$1ms).  $\alpha - \Omega$ dynamo in the first 10s after the
bounce of the core  will generate a  superstrong magnetic field ($B\sim 10^{16}$G).
The  beam of the energy from  the super strongly magnetized milli-second pulsar(SSM-MSP) can punch a hole
in the supernova ejecta . Then the  beam of gamma rays flows out of the
ejecta for $\sim 10$s  with the total energy up to $\sim 2\times
10^{51}$erg. 
If we observe this beam from the lateral
direction,  the energy of gamma rays should be smaller than 500keV
due to the electron scattering  and the total energy
of gamma rays should be much smaller than $\sim 2\times 10^{51}$erg,
which is the very characteristics of NHE-GRBs.
In SSM-MSP model of NHE-GRB, GRB event should be associated with Type Ib/Ic
supernova like GRB980425/SN1998bw.
If  SSM-MSP is produced in a fraction ($\sim$ a few \% ) of Type Ib/Ic supernovae, the event rate as well as the isotropic and homogeneous
distribution of NHE-GRBs are also explained.
In this model  the pulsar with the spin period
$P\sim 1.4s (B/10^{16})(t/{\rm 4month})^{1/2}$ should exist in SN1998bw so that the searches for
this  pulsar in all the wave length are urgent.
SSM-MSPs can accelerate  protons up to $\sim 3\times 10^{22}(P/1{\rm ms})^{-2}
(B/10^{16}$G)eV. If a few \% ($\sim 10^{50}$erg)
of the beam energy of SSM-MSP in NHE-GRB event
 ($> 5\times 10^{51}$erg) is in the form of  high energy protons of energy
$\sim 10^{20}$eV, the flux of the observed UHECRs(Ultra High Energy Cosmic
Rays) can be explained. Then  along the  direction of each UHECR, a
supernova remnant should be found. As a natural consequence of our
model, it is suggested that NHE-GRB is the progenitor of the soft gamma ray
repeater which has also the superstrong magnetic field($B\sim
10^{15}$G) and is in  the supernova remnant.
} \vspace*{0.5cm}\\

\newpage
\section{Introduction}

Galama et al.\cite{galama} reported the discovery of very luminous
Type Ic supernova SN1998bw in the BeppoSAX Wide Field Camera error box 
of  GRB 980425\cite{soffita}, which occurred within about a day of
the gamma ray burst. The duration of GRB 980425 is $\sim$30sec  and 
the burst fluence is $(4.4\pm 0.4)\times 10^{-6}$erg/cm$^2$.
 Since  no burst of emission is detected above 300keV, GRB 980425
belongs to a so called NHE(Non High Energy) burst\cite{pen} 
which is characterized by the lack of fluence above 300keV, 
lower peak flux  and  the 
apparent isotropic and  homogeneous distribution.
A quarter ( $\sim$250 events/y) of GRBs is NHE bursts.\cite{pen}

The host galaxy of SN1998bw is a spiral galaxy at a distance of $\sim$
40Mpc. The chance probability of the coincidence    is  $\sim
10^{-4}$.\cite{galama}  Motivated by the coincidence of Type Ic SN with
GRB, Wang  and Wheeler\cite{wang} examined six recent SN Ib/c for which an
outburst epoch can be estimated with some reliability and found that
all are correlated in time and space with BATSE gamma ray bursts.
Since the joint probability of all six correlations is  $1.5\times  10^{-5}$
 and no such correlation exists for SN Ia and SN II, Wang  and Wheeler
\cite{wang} proposed that all gamma ray bursts are associated
with Type Ib/c supernovae. In this paper I follow the proposal by 
Wang  and Wheeler.  I assume that all NHE-GRBs are 
associated with Type Ib/c supernovae, which should be confirmed in
future observationally. I propose a model that a (C+O) star and
 a neutron star(black hole) close binary is a progenitor of 
NHE gamma ray burst and argue the details of the model. 

In \S 2 a formation scenario of a (C+O) star  and a neutron
star (black hole)  close binary system is discussed. In \S 3 I discuss
a supernova explosion of the (C+O) star which yields a super strongly
magnetized milli-second pulsar(SSM-MSP) . In \S 4 I explain how
SSM-MSP emits NHE-GRB. \S 5 will be devoted to   discussions and
astrophysical implications of our SSM-MSP GRB model.

\section{ Formation of a  (C+O) star  and a neutron star(black hole)  close binary system}
Similar to the scenario of formation of double neutron star systems
\cite{bhat91,yama93}
 we assume that our (C+O) star-neutron star(black hole) binary is formed as
follows\cite{hana95} :
 
1) Consider a binary system composed of two main-sequence stars 1 and 2
of mass $\sim$  25 M$_\odot$ which has a helium core mass
$\sim$ 8M$_\odot$ ,
 
2) Roche lobe overflow of star 1, which becomes a helium star 1,
 
3) the first supernova explosion of helium star 1 to form
a neutron star 1,
 
4) Roche lobe overflow  of star 2 which leads to a spiral-in of
   neutron star 1 into star 2 . The system shrinks and finally
it consists of a neutron star 1 and a helium star 2.
Here we note that if the super Eddington accretion is
possible\cite{chev93} a neutron star 1 becomes a black hole of mass $\sim
2.4\sol$ in this spiral-in phase\cite{bethe98}.

5)  During the presupernova evolution of   helium star 2, the helium
envelope expands as its core contracts. The radii of star 2 with mass
$3\sim 8\sol$  are $4\sim 1R_\odot$ \cite{nomoto88,yama93}. 
So if the Roche  radius of
star 2 is smaller than $4\sim 1R_\odot$ , a spiral-in of neutron
star(black hole) 1 occurs again. Now the system consists of (C+O) star 2 and neutron star
(black hole) 1. The mass of the (C+O) cores are 6.0 and 3.8 M$_\odot$
 for mass of helium stars 8 and 6M$_\odot$,
 respectively \cite{nomoto88}, while the radius of
the  (C+O) star is $\sim$ 0.1R$_\odot$.
Since the radii of helium stars of mass 8 and 6$\sol$
are 1.3 and 1.9 R$_\odot$, respectively,
 at the onset of Roche lobe overflow,
 the separation after the spiral-in of neutron star(black hole) 1 will
be significantly smaller than 1R$_\odot$, say, 0.2R$_\odot$.
Such a  (C+O) star in a close binary may be a progenitor of Type Ic
supernova such as SN1994I\cite{nomoto94}.

\section{ Super Strongly Magnetized Milli-Second Pulsar}
Now I discuss an evolution of our  close binary system with separation
$a \sim$ 0.2 R$_\odot$ which consists of
 a (C+O) star of mass $M_2\sim$ 6 M$_\odot$ and a
neutron star(black hole) of mass $M_1=1.4(2.4)$M$_\odot$
\cite{hana95}. There are
three important time scales in this system:
 
1) $t_{ev}$ : the evolution time of the (C+O) star. They are
100 yr and 450yr for $M_2=6$ and 3.8
M$_\odot$, respectively\cite{nomoto88}.
 
2) $t_{gw}$ :coalescing time of the binary due to the emission of
gravitational waves which is given by
\bea
t_{gw} = 4\times 10^3 yr(\frac{M_1}{1.4M_\odot})^{-1}
(\frac{M_2}{6M_\odot})^{-1}(\frac{M_1+M_2}{7.4M_\odot})^{-1}
(\frac{a}{1.4\times 10^{10}cm})^4 .
\eea
 
3) $t_{synch}$ :the tidal synchronization time  which is given by\cite{bild92}
\bea
t_{synch}\sim \frac{I\Omega_{orb}}{N},
\eea
\bea
\Omega_{orb} = 1.9\times 10^{-2} s^{-1}(\frac{M_1+M_2}{7.4M_\odot})^{0.5}
(\frac{a}{1.4\times 10^{10}cm})^{-1.5},
\eea
where $I, \Omega_{orb}$ and {\it N} are the moment of the inertia
of the (C+O) star 2, the orbital angular velocity and the tidal
torque, respectively. {\it N} is estimated as
\bea
N = \frac{G M_1^2}{R} (\frac{R}{a})^6\alpha ,
\eea
where R and $\alpha$ are the radius of star 2 and the angle between
the orbital separation vector and the tidal bulge, respectively.
 $\alpha$ is expressed by the viscous damping time $t_{vis}$
 as
\bea
\alpha = \frac{\Omega_{orb} R^3}{G M_2 t_{vis}}  .
\eea
Therefore $t_{synch}$ becomes
\bea
t_{synch} = (\frac{M_2}{M_1})^2 (\frac{a}{R})^6 t_{vis} \sim
              1800 t_{vis} .
\eea
Since the convective core exists in the (C+O) star\cite{nomoto88}, 
$t_{vis}$ is estimated as
\bea
t_{vis}= (\frac{R}{v_{conv}})( \frac {R}{l_m})
\eea
where $v_{conv}$ and $l_m$ are velocity of convection and the mixing
length, respectively. Putting $v_{conv} \sim 10^5$cm s$^{-1}$ and
 $l_m \sim 0.2 R$ , we have
\bea
t_{synch}\sim 20yr(\frac{v_{conv}}{10^5cm s^{-1}})^{-1}
          (\frac{l_m}{0.2 R})^{-1}\frac{R}{7\times 10^9cm}
\eea
 
Since we have the inequality as
\bea
 t_{gw}\geq t_{ev} \geq t_{synch},
\eea
the (C+O) star 2 is tidally locked in the carbon burning
phase  and the supernova
explosion of star 2 occurs with essentially the same separation
{\it a}.  In tidally locked binary the spin angular
velocity is the same as the orbital one so that  the collapsing central
1.4M$_\odot$ core of star 2 will have  the spin angular momentum {\it J} estimated as
\bea
J = I_{core} \Omega_{orb},
\eea
where $I_{core}$ is the moment of inertia of the  central
1.4M$_\odot$ core.  Inserting $I_{core}$ of  $10^{50}\sim 10^{51}$gcm$^2$ in the carbon
 burning phase\cite{nomoto88}, we have
\bea
J=2\times 10^{48\sim 49}{\rm gcm}^2 {\rm s}^{-1}(\frac{M_1+M_2}{7.4M_\odot})^{0.5}
(\frac{a}{1.4\times 10^{10}cm})^{-1.5}.
\eea
This angular momentum is  comparable to that of 1 ms
pulsar ($\sim 6\times 10^{48}$g cm$^2$s$^{-1}$),
 which means that the new born pulsar after the supernova explosion
of star 2 is necessarily a milli-second pulsar.
 
Duncan and Thompson\cite{dun92,dun93} argued 
$\alpha - \Omega$ dynamo theory for such a
rapidly rotating new born pulsar and concluded that such a milli
second pulsar may have a superstrong magnetic field of $\sim
10^{16}$G within 10s after the formation. Although Duncan and Thompson 
considered the accretion
induced collapse, the same argument can apply to our (C+O) star.
Recently  such a superstrong magnetic field ($\sim 10^{15}$G)
has been identified in a soft gamma ray repeater SGR1806-20\cite{sgr}
with  the age $\sim 10^4$y. This observation supports a model of soft gamma 
ray repeaters based on super strongly magnetized  neutron stars
\cite{dun95,dun96}. Therefore even relying only on the
observational results, it may be possible  to consider a superstrong
magnetic field as strong as $\sim 10^{16}$G in a new born milli-second pulsar. 
Considering both theoretical suggestions and observations of such 
a superstrong magnetic field, we assume here that after the
explosion of our (C+O) star a milli-second pulsar with  the superstrong magnetic field ($B\sim
10^{16}$G) is formed in $\sim 10$ s. 
Here  note that in vacuum there exists the maximum electric field ($E_c =
2\pi m_e^2c^3/he=4.4\times 10^{13}$e.s.u) while the maximum magnetic field does
not exist  since the  vacuum with such a
strong field is shown to be unstable only for $(E^2-B^2)>0$. \cite{vac} 
(See page 677 of reference 18.)

\section{Supernova explosion of a (C+O) star and NHE gamma ray bursts}  
Now star 2 explodes as a Type Ic supernova  
with the ejecta of mass $M_e=2\sim 5\sol$. The column density
($\Sigma$) of the ejecta at time $t$ from the explosion is given by
\be
\Sigma\sim 10^5{\rm
g/cm}^2(\frac{M_e}{\sol})^2E_{51}^{-1}(\frac{t}{{\rm 1day}})^{-2}
\ee
where $E_{51}$ is the explosion energy in the unit of $10^{51}$erg.
Recently Woosley, Eastman and Schmidt\cite{woos98} made light curve models for
SN1998bw and found that a
6$\sol$ (C+O) star model with the unusually large explosion energy of $\sim 2\times
10^{52}$ erg can fit the light curve quite well. They also found that
both shock break-out and relativistic shock deceleration in
circumstellar material fail to produce a GRB of even the low
luminosity($\sim 10^{48}$erg) inferred for GRB980425.
This means that the source of gamma rays should be inside the ejecta.
Since the time difference between the supernova explosion and GRB
980425 is one day or so, from Eq. (12) the column density at $t\sim$
1day is so high  that any gamma
rays  emitted from the central new pulsar, neutron star or black
hole can not reach the surface of the ejecta unless there is  a hole
 in the ejecta.  This reminds us the mysterious spot in
SN1987A\cite{mys1,mys2}  which appeared 5-7 weeks after the explosion
of Type II SN1987A at $\sim$ 0.06 arc-sec away from SN1987A. 

To explain mysterious spot in SN1987A, at least  four mechanisms 
were examined to punch a hole in the ejecta\cite{rees87,piran87}.
These mechanisms can apply to SN1998bw/GRB980425.
They are;
\begin{enumerate}
\item Due to the existence of the secondary (a neutron star or black hole
in this case) the density in the direction shadowed by the companion
will be low\cite{rees87}.

\item Rayleigh-Taylor instability in the ejecta may make a hole\cite{rees87}.

\item The relativistic jets formed at  the core collapse punch
 holes in the ejecta\cite{piran87}.

\item Beams of energy  from the super strongly magnetized milli
second pulsar may punch holes in the ejecta\cite{piran87}.
\end{enumerate}

In the first mechanism the relativistic plasma from the pulsar
gradually inflates a bubble within the remnant. Since the remnant is 
punctured, the relativistic plasma from the over-pressured bubble would
squirt into the surrounding. However the time needed for this
mechanism to operate would be much longer than $\sim$ 30s unless
 the direction of the hole  coincides with
the direction of the beam from the pulsar or from the black hole with
the accretion disk. In the second mechanism, the chance probability of such a hole would be
very  low for t$\sim$30 s.  A chance coincidence of
the hole and the direction of the beam is also needed. 

In the third mechanism the direction
of the hole should agree with the direction of the spin of the pulsar 
or the rotating axis of the black hole. General relativistic numerical
simulations\cite{naka81} have demonstrated that a rapidly rotating 
core collapses first along its rotation axis. Collapse continues along
the equatorial directions, then bounces along the rotation axis to
form jets. In one example, a 1.4$\sol$ stellar core modeled as a
$\gamma=3/4$ polytrope produced jets with mass 0.007$\sol$, kinetic
energy $5\times 10^{51}$ erg and average velocity 0.6c. Since the
opening angle of this jet is a few degree, it is possible to punch a
hole at the first 10 seconds or so. The total angular momentum
in this model  $2.4 \times 10^{49}$gcm$^2$s$^{-1}$ is four times larger
than 1ms pulsar. Although this kind of jets may also be formed 
in MHD rotating core collapses \cite{leblanc70}
, in general,  with the increase of the angular momentum
the kinetic energy of the explosion decreases since the bounces occur
at the  lower density and weaker gravitational potential, while SN1998bw has
the unusually large kinetic energy. 
We therefore consider the fourth mechanism mainly in this paper.

 For the typical angular momentum
in the core of our (C+O) star corresponding to a milli-second pulsar, $\sim 6\times 10^{48}$gcm$^2$/s, the collapse is
essentially spherically symmetric because the ratio of the centrifugal
force to the gravity is $\sim$0.2 even for the neutron star radius.
A milli-second pulsar has the rotational energy($E_{rot}$) given by ,
\be
E_{rot}=2\times 10^{52}{\rm erg} I_{45}P_{ms}^{-2},
\ee
where $I_{45}$ and $P_{ms}$ are the moment of inertia in the unit of $10^{45}$gcm$^2$
 and the spin period in milli-second, respectively.
The spin period  and the
luminosity ($L$) of the pulsar are given by 
\ba
P_{ms}&=& \sqrt{(\frac{P_i}{1ms})^2+\frac{t}{\tau}}, \\
\tau &=& 5{\rm s}B_{16}^{-2}I_{45}R_6^{-6}, \\
L &=&3.8\times 10^{51}{\rm erg/s} B_{16}^{2}R_6^{6}((\frac{P_i}{1ms})^2+\frac{t}{\tau})^{-2}, 
\ea
where $P_i$ , $ B_{16}$ , $R_6$ and $t$ are the initial spin period,  the magnetic field in unit of $10^{16}$G
, the radius of the neutron star in unit of $10^6$ cm and the time
after the bounce of the core, respectively.

The expansion velocity ($V_{exp}$) of the ejecta is given by
\be 
V_{exp}\sim 10^9{\rm cm/s} (\frac{M_e}{\sol})^{-1/2}E_{51}^{1/2}.
\ee
For SN1998bw the observed expansion velocity is $\sim 2\times 
10^9$cm/s\cite{vel}. This means that  $E_{51}\sim 
20$  for $M_e \sim 5\sol$. This unusually large explosion energy is consistent with a
model of light curves\cite{woos98}.
The shock will reach the surface of the (C+O) star($\sim 10^{10}$cm) 
in $\sim 5$s. Since the magnetic field will be amplified to $\sim 10^{16}$G
within 10s\cite{dun92}, the superstrong magnetic field pulsar begins
to operate after or just when the shock breaks out. The solid angle
extended by the beam from the pulsar is given by\cite{gold}
\be
\Delta \Omega =\pi \frac{2\pi R}{Pc}\sim 0.05\times 4\pi P_{ms}^{-1}
\ee 
Then the  energy of $\sim 10^{52}$ erg would be injected into 5\% of the
ejecta in $\sim 10$s. If this energy is converted to the kinetic
energy of the ejecta within the cone of the beam, the velocity
($V_{cone}$) 
would be
\be
V_{cone}=1.4\times 10^{10}{\rm cm/s} (\frac{M_e}{\sol})^{-1/2}
\ee
This is much larger than $V_{exp}$ so that a hole is punched out in
the ejecta. 

 If, however, $E_{51}\sim 1$ and $M_e \sim 5\sol$ as usual
, $V_{exp}\sim 5\times 10^8$cm/s and the shock will reach the surface 
in $\sim 20$s so that the pulsar activity would start
before the shock breaks out. Then  a part of the 
 energy of the beam would be used to expand the ejecta as a whole so that an 
effective explosion energy would increase and we may have 
$V_{exp}\sim 10^9$cm/s finally.
 In this case $V_{cone}$ will
decrease but as far as $V_{cone} >  V_{exp}\sim 2\times 10^9$cm/s 
, we may regard the hole is punched out. For this the minimum energy 
needed is $\sim 2\times 10^{51}$ erg so that almost all the
rotational energy of the milli-second pulsar might  be used to 
the expansion energy of  the ejecta.
Note here that the unusually large explosion energy of SN1998bw is the
same as the rotational energy of the milli-second pulsar.
This may not be  a chance coincidence.
In any case a hole would be punched out by the beam of the pulsar in
the first 10 s or so.

We assume here that at $t\sim 10$s the hole is punched out.
Then  the period of the pulsar is $\sim 2$ms and the rotational
energy is $\sim 5\times 10^{51}$erg.  The energy 
beam of the pulsar will flow out from the hole. This
situation is similar to the millisecond pulsar models of GRBs 
based on the accretion induced collapse of magnetized white dwarfs
\cite{usov92,usov94}. 
The beam of the energy would be finally converted to the
beam of  gamma rays \cite{usov92,usov94}. As a whole 
$\sim 5\times 10^{51}$erg of gamma rays would be emitted in a 
 cone  of the solid angle $\sim 0.025\times 4\pi$. 
The chance probability that the line of the sight is in the cone is
$\sim 0.025$. 
We will observe the beam of the gamma rays, in general, 
from the lateral direction so that only gamma rays 
scattered by the thin matter blown off from the ejecta  in the first
10s will come along the line of sight .

  In  Compton scattering  the energy of the scattered photon
($\omega$)  is given by  
\be
\frac{1}{\omega}=\frac{1}{\omega_0}+\frac{1-\cos \theta}{m_ec^2}
\ee
where $\theta$ is the scattered angle and $\omega_0$ is the incident
energy. Even for $\omega_0 \gg m_ec^2$,
unless  $\theta$ is small such that $\omega_0(1-\cos \theta) \ll
m_ec^2$, $\omega < m_ec^2$. That is , the energy of the scattered gamma ray should 
be smaller than 500keV. This explains  the lack of fluence above
300keV for NHE-GRBs. The scattered photon is more or less
distributed isotropically, that is, it is observed from almost all the 
direction and the total energy would be much smaller than $\sim 10^{51}$
erg because only a small fraction  of the beamed gamma rays  would be
scattered by  the very low density plasma. 
 This is also consistent with the total energy ($\sim
10^{48}$erg) of GRB980425.

About a fourth of gamma ray bursts($\sim$250
events/y) is NHE-GRBs \cite{pen} . The event like
GRB980425 will not be observed if it occurred at 20 times farther $\sim$ 800 Mpc. The
number of galaxies within 800Mpc is $\sim 2\times 10^7$ and the
event rate of TypeIb/Ic  supernova is $\sim 10^{-3}$/y/galaxy\cite{capp}
 so that the event rate of 
TypeIb/Ic supernova within 800Mpc is $\sim2\times 10^4$/y. This suggests
that only a fraction ($\sim$ a few\%) of TypeIb/Ic  supernova becomes
NHE gamma ray burst which is different from the suggestion by Wang and
Wheeler.\cite{wang} This is , however, consistent with our model. 
Depending on the initial separation
and initial mass of the binary, only a fraction of the binary yields
the super strongly magnetized milli-second pulsar.

After an NHE-GRB event the luminosity of the pulsar is given by
\be
L=1.3\times 10^{43}{\rm erg/s}
B_{16}^{-2}I_{45}^2R_6^{-6}(\frac{t}{{\rm 1day}})^{-2}
\ee 
For SN1998bw, the above luminosity is much smaller than the observed
luminosity so that the existence of the pulsar will  not affect the light
curve of the supernova.

\section{Discussion}

It is not clear at present that our model can explain 
 HE(High Energy) GRBs and its afterglow also.
However if we observe NHE-GRB in our model along the beam direction
and if we regard that the energy is emitted isotropically, then  the total
energy amounts to $\sim 2\times 10^{53}$erg,
which is comparable to the maximum observed energy of GRBs\cite{kul98}.
 We here argue the event rate of NHE-GRBs up to $z\sim 3$.
Adopting that $\Omega_M\sim 0.2$ and $\Omega_{\Lambda}\sim 0.8$ and
$H_0=$65km/s/Mpc, the total number of NHE-GRBs up to $z\sim 3$
 is $\sim 10^5$/y. If NHE-GRBs observed along the cone of the pulsar
beam look like HE GRBs, the event rate of HE GRBs would be $\sim 10^3$/y,
which is comparable  to the event rate of HE GRBs. 

The maximum energy of a proton accelerated by the SSM-MSP is given by
\cite{gold}
\be
E_{max}=7.5\times 10^{21}eV R_6^3B_{16}(\frac{P}{2ms})^{-2}.
\ee
This is larger or comparable to the maximum energy of the
cosmic rays so far observed. Such a high energy cosmic ray can 
propagate only up to the distance of $\sim$ 100Mpc due to the
production of pion through the collision with the microwave 
background radiation. Then the number density of  UHECRs(Ultra High Energy
Cosmic Rays $ E >10^{20}$eV) is estimated as $n_{cr}\sim
10^{-30}$cm$^{-3}$. \cite{wax95}   If UHECRs are supplied by NHE-GRBs
, the total energy of UHECR per one NHE-GRB-event is estimated as
$\sim 10^{50}$erg.  This energy is only 0.05\% of the total energy of the milli-second pulsar 
and 2\% of the energy emitted by the pulsar after the hole is punched
out in our model. Very recently Agasa\cite{agasa} team reported 
possible clustering in position of UHECRs, which may be in favor of
GRB model of  UHECRs\cite{wax96}. Since the time broadening of the pulse
of UHECRs due to the intergalactic magnetic field is more than 100 y
\cite{wax95}, the arrival time of UHECR does not coincide with the GRB
event. However in our model there should exist the supernova remnant
in  the direction of each UHECR if the time broadening is shorter than
the maximum age of the observable supernova remnant.

The relation between soft gamma ray repeaters (SGRs) and NHE-GRBs in our
model is quite interesting.  NHE-GRBs in our model is a strongly 
magnetized pulsar in the supernova remnant with the event rate of
$10^{-4}\sim 10^{-5}$events/y/galaxy. This is exactly the same as the 
characteristics of  soft gamma ray repeaters. This suggests that
progenitors of SGRs are NHE-GRBs in our model.

\section*{Acknowledgments}
The author would like to thank Prof. F. Takahara for useful discussions.
This work was supported in part   by
Grant-in-Aid of Scientific Research of the Ministry of Education,
Culture, Science and Sports, 09640351.

\newpage

\end{document}